\documentstyle[epsfig]{ioplppt}

\begin{document} 
\jl{01}

\title{Sluggish Kinetics in the Parking Lot Model}

\author{J. Talbot\dag, G. Tarjus\ddag, and P. Viot\ddag} \address{\dag
Department   of  Chemistry   and  Biochemistry, Duquesne   University,
Pittsburgh, PA 15282-1530}   \address{\ddag  Laboratoire  de  Physique
Th\'eorique des Liquides, Universit\'e Pierre et Marie Curie, 4, place
Jussieu 75252 Paris, Cedex, 05 France}
\begin{abstract}
We  investigate, both  analytically and  by   computer simulation, the
kinetics of a  microscopic model of  hard rods  adsorbing on a  linear
substrate.   For   a small, but  finite   desorption rate,  the system
reaches the equilibrium state  very slowly, and the long-time kinetics
display  three successive regimes: an algebraic  one where the density
varies   as $1/t$, a logarithmic    one where the   density varies  as
$1/ln(t)$, followed by a terminal  exponential approach.  A mean-field
approach  fails to  predict the relaxation   rate associated with  the
latter.  We show that the correct answer can only be provided by using
a systematic description based on a gap-distribution approach.
\end{abstract}

In many situations,  thermal energies  are significantly smaller  than
the energy needed  for  hard particles  to diffuse, and  under applied
external   forces  the    system   evolves towards     non-equilibrium
configurations,  metastable  configurations, or  eventually  reaches a
stable equilibrium state after a very  slow process.  For example, the
adsorption of   some  proteins and    colloidal particles   on   solid
surfaces\cite{ramsden} involves particle-surface  energies that are  so
strong that the process is characterized by extremely small desorption
and  surface diffusion on   the experimental time scale.   In granular
materials, particles are trapped in a metastable configuration, unless
external    energy    is   brought    to     the     system.    Recent
experiments\cite{knight95,nowak98} measured   the densification  of  a
vibrated granular   material.    A  column  containing    monodisperse
spherical beads was tapped periodically with a given intensity and the
powder evolved slowly, essentially as the  inverse of the logarithm of
the number of taps,  from a loosely  packed  state to a  denser steady
state   whose density depends  on  the tapping strength.  In all these
cases,   geometric exclusion    effects dictate   the    kinetics   of
densification, i.e.,  addition  of   new particles is    exponentially
limited by the  inverse  of the free  volume\cite{nowak98,boutreux97}.
These   effects are accounted  for  in  a simple adsorption-desorption
model\cite{nowak98,jin94,krapivsky94}.   Its  one-dimensional version,
also known as  the parking lot model,  was  shown to display a  $1/\ln
(t)$       approach to the  final   state     for vanishing desorption
rate\cite{jin94,krapivsky94}.  More    recently,   Ben  Naim  {\em  et
al.}\cite{bennaim98} proposed an approximate solution of the model for
a small  but non-zero desorption rate.  We  discuss  here the solution
and show that the  mean-field  treatment (or adiabatic  approximation)
used in the   late  stage of  the  process  is not  valid  and greatly
underestimates the characteristic  time of  relaxation.  We develop  a
more consistent approximation  which   shows good agreement  with  the
numerical simulations.

The general  definition of   the  adsorption-desorption model   is  as
follows.  Attempts are made  to add objects  to  a space of  arbitrary
dimension at randomly selected positions with a constant rate $k_{+}$.
If the trial position does not result in  an overlap with a previously
placed  object, the new object  is accepted.  In addition, all objects
in the system are  subject to  removal (desorption)  at random  with a
constant rate $k_{-}$.  In the  parking lot model,  the substrate is a
line and the objects are hard rods.  This  $1$-d model has been solved
in  some  limiting cases.  When $k_{-}=0$,   the adsorption is totally
irreversible and the process corresponds  to a $1$-d Random Sequential
Adsorption    (RSA)  for  which   the   kinetics   are known   exactly
\cite{evans93}.   Due to the absence    of relaxation mechanism,  this
process  evolves towards  a  non-equilibrium state  and the  long-time
kinetics     are    given     by    an      algebraic    scaling  law,
$\rho_{\infty}-\rho(t)\sim 1/t$, with    $\rho_\infty\simeq  0.747..$.
When $k_{+}=0$, one recovers a desorption process for which analytical
solutions   are   also available\cite{VanTassel97}.    In    the limit
$k_{-}\rightarrow0$,   accurate   descriptions     have           been
obtained\cite{jin94,krapivsky94}.   In  this case, the process cleanly
divides into two sub-processes.   The   initial phase consists of   an
irreversible adsorption and it is  followed by an infinite sequence of
desorption-adsorption events in which  a rod detaches from the surface
and the gap  that is created is immediately  filled by one  or two new
rods.  The latter possibility causes the system to evolve continuously
to the close-packed state with $\rho=1$, as \cite{jin94,krapivsky94} $
1-\rho(t) \simeq 1/\ln(t\ln(t)) $ where $t$  now represents a rescaled
time.  For the general case, where both $k_+$  and $k_-$ are non zero,
no complete solution is available.

\begin{figure}
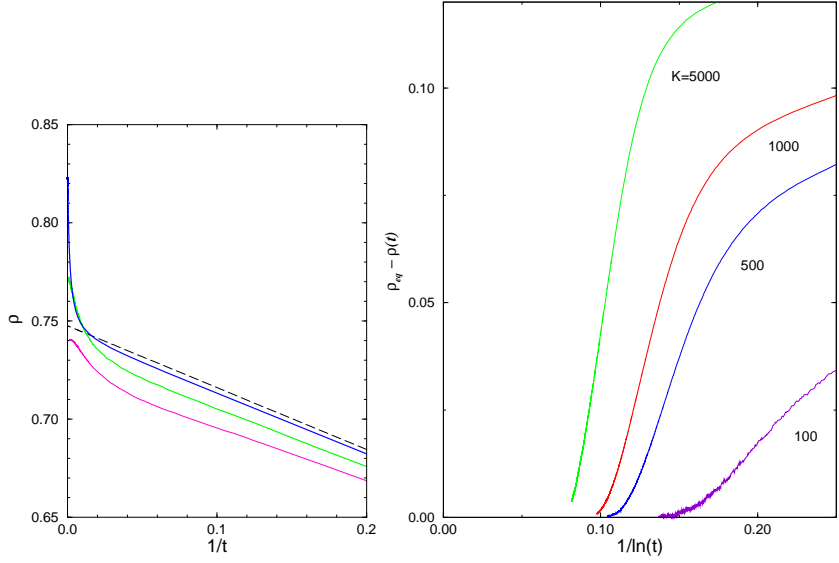

\begin{center}
\resizebox{12cm}{!}{\includegraphics{figure1.pstex}
\includegraphics{figure2.pstex}}
\caption{(a)   Evolution  of  the adsorbed  density  versus  $1/t$ for
several values of $K$, ($K=100, 500, 5000$ from bottom to top) and for
a  RSA  process ($K=\infty$, dashed   curve).    (b) Evolution of  the
adsorbed density to its equilibrium value.  At intermediate times, the
adsorbed  density evolves   as  $1/\ln(t)$,   while   very close    to
equilibrium the density relaxes exponentially.}
\label{fig:rho}
\end{center}
\end{figure}

\begin{figure}
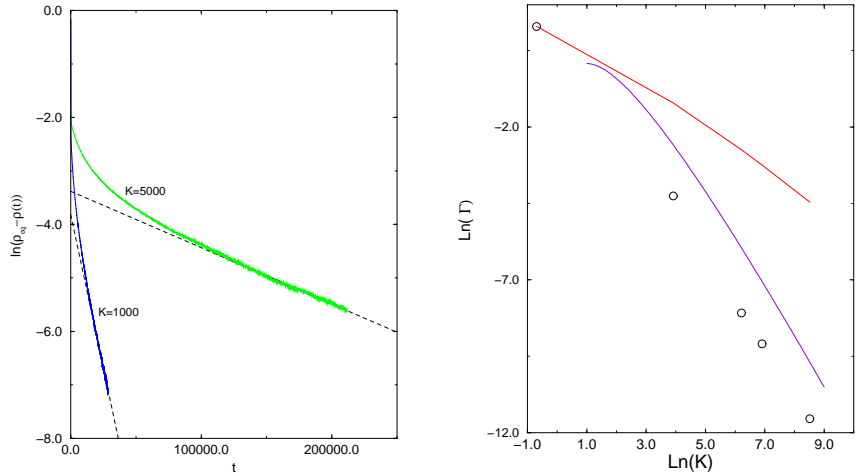

\begin{center}
\resizebox{12cm}{!}{\includegraphics{figure3.pstex}
\includegraphics{figure4.pstex}}
\caption{(a)    Final  exponential  approach  of the    density to its
equilibrium value. (b) Relaxation rate $\Gamma$ for the approach to
equilibrium. Upper  curve:  prediction from mean-field  approximation,
Eq.~(\ref{aeq}).  Intermediate curve: prediction from Eq.~(\ref{aex}).
Open circles: from the numerical simulations.}\label{fig:rate}
\end{center}
\end{figure}

The properties of the parking lot model depend only on  the ratio $K =
k_{+}/k_{-}$.  With an appropriate rescaled time, the kinetic equation
describing the evolution of the density of adsorbed particles is given
by
\begin{equation}
\frac{d\rho}{dt} = \Phi(t)-\frac{\rho}{K},
\label{kinet}
\end{equation}
where   $\Phi$, the insertion  probability  at  time  $t$  (or density
$\rho$),  is the fraction of the  substrate  that is available for the
insertion of a new particle.  Thus,  large values of $K$ correspond to
small desorption rates.  The presence of  a relaxation mechanism, even
infinitesimally  small,  implies that the   system   eventually reaches a
steady state that corresponds  to an equilibrium configuration of hard
particles with $\rho_{\rm  eq} = K\Phi_{\rm eq}(\rho_{\rm eq})$, where
$\rho_{\rm eq}$ denotes  the equilibrium density.  At equilibrium, the
insertion    probability   is given exactly by
\begin{equation}
\Phi_{\rm eq}(\rho)= (1-\rho)\exp(-\rho/(1-\rho)).
\label{phieq}
\end{equation}
Inserting   Eq.~(\ref{phieq})  in  Eq.~(\ref{kinet})   leads  to   the
following expression for the equilibrium density:
\begin{equation}
\rho_{\rm eq} = \frac{L_w(K)}{1+L_w(K)}
\label{isotherm}
\end{equation}
where  $L_w(x)$ (the Lambert-W  function) is the solution of $x=ye^y$.
In  the limit  of   small $K$ the isotherm    takes the Langmuir  form
($\rho_{eq}\sim   K/(1+K)$)   while for    large   $K$, $\rho_{eq}\sim
1-1/\ln(K)$.  At small values of $K$, equilibrium is rapidly obtained,
but at  large values  the  densification is   extremely slow.  In  the
following, we restrict ourselves  to the latter  case.  In the initial
stages of the  process,  desorption events are negligible  compared to
adsorption   and   the  process   displays   an   RSA-like   behavior.
\Fref{fig:rho}a shows  simulation results (see  below for details) for
the  density versus     $1/t$   in the intermediate    density  region
($0.4<\rho(t)<0.75$).  The  larger  the value  of $K$,  the larger the
scaling law region.    For $\rho(t)>0.7$, which   implies $K\geq 100$,
desorption can no   longer be ignored,   because it  permits  particle
rearrangements on  the line and,  eventually,  insertion of additional
particles.  The mechanism for densification is  similar to that of the
model with infinitely small desorption  and the kinetics is  dominated
by the $1/\ln(t)$ term (\Fref{fig:rho}b).  For large but finite values
of  $K$, this regime  persists until the density is  very close to the
equilibrium value  and the    desorption term  is comparable  to   the
adsorption term.  In  the final regime,  an  exponential approach to
equilibrium  is observed (\Fref{fig:rate}a).   This  terminal approach
corresponds to the linear response regime.

To determine the kinetics of the densification process, a knowledge of
$\Phi(\rho)$ is required.  The evolution equations can be expressed by
means of the gap distribution functions  $G(h,t)$ which represents the
density of  voids of length $h$.   The time derivative of  $G(h,t)$ is
given by
\begin{eqnarray}
\frac{\partial{G(h,t)}}{\partial    t}&=&   -H(h-1)(h-1)  G(h,t)   + 2
\int_{h+1}^{\infty} dh'   G(h',t)\nonumber\\      &&-\frac{2}{K}G(h,t)
+\frac{H(h-1)}{K\rho(t)}           \int^{h-1}_{0}                  dh'
G(h',h-1-h',t),\label{eqcin1}
\end{eqnarray}
where $G(h,h',t)$  is the gap-gap  distribution function and $H(x)$ is
the unit step function. The first two terms on  the right-hand side of
Eq.~(\ref{eqcin1}) correspond to  the   loss and  gain terms  due   to
adsorption while the  remaining two are due  to desorption.  A similar
equation can be written for $G(h,h',t)$,  and it requires higher-order
gap   distribution   functions.   As expected,    even   though  it is
one-dimensional, this kind of  process leads to an infinite  hierarchy
of equations involving   an  infinite set of   multi-gap  distribution
functions.  Only approximate solutions can  be  obtained, by means  of
closure ansatz that enables the truncation of the hierarchy.

\begin{figure}
\begin{center}
\resizebox{11cm}{!}{\includegraphics{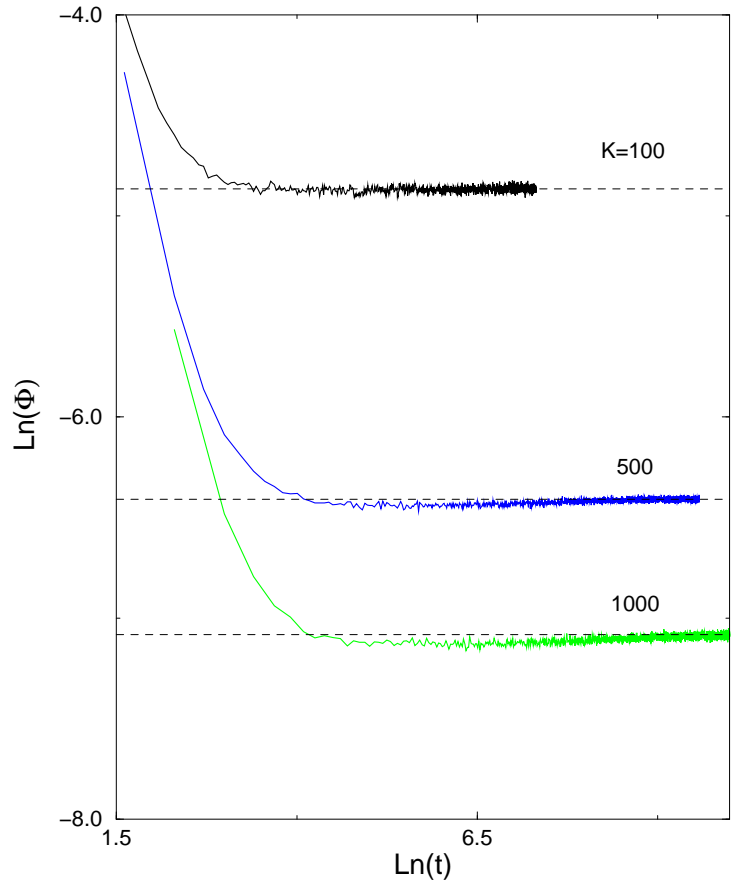}} 
\caption{Approach of   the insertion  probability  to  its equilibrium
value       given   by    Eq.~(\ref{phieq})   as     a    function  of
$\ln(t)$.}
\label{fig:phi}
\end{center}
\end{figure}

The   insertion  probability can  be  expressed in   terms  of the gap
distribution function as
\begin{equation}\label{phi}
\Phi(t)=\int_1^\infty dh (h-1) G(h,t),
\end{equation}
and we have in addition the following sum rules:
\begin{equation}\label{rho}
\rho (t)=\int_0^\infty dh G(h,t)= 1- \int_0^\infty dh\,h G(h,t).
\end{equation}
The  exact    expression  for   the  equilibrium structure    is known
\cite{jin94,krapivsky94};   the gap distribution function is 
\begin{equation}
G_{\rm               eq}(h,\rho)     =     \frac{\rho^2}      {1-\rho}
\exp\biggr(-\frac{\rho}{1-\rho}h\biggr),
\label{geq}
\end{equation} 
and all  higher-order distribution functions satisfy the factorization
property,
\begin{equation}
G_{\rm   eq}(h_1,h_2,\cdots,h_n,\rho)=    G_{\rm   eq}(h_1,\rho)G_{\rm
eq}(h_2,\rho) \cdots G_{\rm eq}(h_n,\rho).
\label{factor}
\end{equation}
An  adiabatic (mean-field) approximation assumes  that, at any density
$\rho(t)$, the structure of   the adsorbate acquires very   rapidly an
equilibrium form satisfying Eqs.~(\ref{geq}) and~(\ref{factor}).  This
leads  to an expression  for   $\Phi$ akin  to  Eq.~(\ref{phieq}) with
$\rho(t)$     in   place  of     $\rho_{eq}$.     Therefore, expanding
Eq.~(\ref{kinet}) to first order in density, $\delta \rho(t) = \rho(t)
- \rho_{\infty}$, with $\rho_{\infty}= \rho_{eq}(K)$, one obtains
\begin{equation}
\frac{d}{dt}\delta\rho = -\Gamma_{MF}(K)\delta\rho + O(\delta\rho^2)
\label{expand}
\end{equation}
with
\begin{equation}
\Gamma_{MF}(K) = \frac{(1+L_w(K))^2}{K},
\label{aeq}
\end{equation}
which implies an exponential approach to  the equilibrium state with a
relaxation time given  by $K/\ln(K)^2$  for  large $K$.  Using  a very
efficient   algorithm     for  the   adsorption-desorption   processes
\cite{talbot99},  we   have performed  Monte-Carlo   simulations for K
ranging from  $10$ to  $5000$.    Averages  were taken  over    $5000$
independent runs.  \Fref{fig:rate}b  shows  the relaxation  rate versus
$K$: the full  curve gives the mean-field prediction, Eq.~(\ref{aeq}),
and dots correspond  to simulation  results.   It is evident  that the
mean-field  analysis gives a poor estimate  of the relaxation rate for
large  $K$.    In   contrast  with  adsorption    models   with   fast
diffusion\cite{privman92}, the mean-field  approximation is  not valid
here; the desorption mechanism  becomes so  inefficient for large  $K$
that  the system does   not follow a quasi-static   path even for  the
regression of small  fluctuations to equilibrium.   In \Fref{fig:phi},
one   observes that  the insertion  probability  $\Phi(t)$ displays  a
non-monotonic behavior at long  times for $K>100$, and {\em approaches
the  equilibrium value from  below}.   Hence, at  long times, the time
derivative  of $\Phi$  becomes  positive.  As the  density remains  an
increasing  function of  time along the  entire   process, the density
derivative of the  insertion  probability is positive and   bounded by
$1/K$ by  virtue of Eq.~(\ref{kinet}).   Consequently, this shows also
that the  relaxation rate is smaller  than $1/K$.  (Let us recall that
both the density and time derivatives of $\Phi$ are always negative in
the adiabatic approximation.)

To  obtain  the   leading term  in the    exponential approach towards
equilibrium, when  $K$ is very large (but  finite), we assume that, as
for the steady state, $|G(h,t)|\sim \exp (-\Pi h)$, with $\Pi \sim \ln
K$. As  a consequence, if one defines $\rho_n(t)=\int_n^{n+1}G(h,t)dh$
and   $\Phi_n(t)=\int_n^{n+1}(h-1)G(h,t)dh$,     then      $\rho_n\sim
\Phi_{n-1}\sim K^{-n}$, so that if one looks for the dominant behavior
in $1/K$, it is sufficient to consider  the first intervals in $h$. As
in  the adiabatic approximation, one  can  expand the gap densities in
power of $\delta\rho(t)$  and keep  only the  linear  term which gives
rise to the exponential decay.  More specifically, by introducing
\begin{equation}\label{defah}
A(h)=\rho_{\infty}\left.\frac{\partial      \ln    G(h,\rho)}{\partial
\rho}\right|_{\rho_\infty}
\end{equation}
where  $A(h)$ is a  piecewise  continuous function, Eq.~(\ref{eqcin1})
can be rewritten for $0<h<1$ as
\begin{equation}\label{defgam}
\frac{2-\gamma}{K}A(h)=\frac{2P_\infty}{K}
\int_h^{+\infty}dh'e^{-P_\infty(h'-h)}A(1+h')
\end{equation}
where    $P_\infty=\rho_\infty/(1-\rho_\infty)$ is   the   equilibrium
pressure for   $\rho_\infty  =   \rho_{eq}$  and  $\gamma=   -K\Gamma=
-K\delta\dot{\rho(t)}/        \delta\rho(t)|_{\rho_\infty}$.      From
Eqs.~(\ref{kinet}) and~(\ref{phi}), one obtains
\begin{equation}\label{gameq}
\gamma=1 -P^2_\infty\int_0^\infty dh h e^{-P_\infty h} A(1+h)
\end{equation}
whereas the sum rules in Eq.~(\ref{rho}) give
\begin{equation}\label{rules}
P_\infty\int_0^{\infty}   dh e^{-P_\infty   h}A(h) =         -P_\infty
\int_0^{\infty} dh h e^{-P_\infty h} A(h)=1.
\end{equation}
When  integrating the two  sides of  Eq.~(\ref{rules}) between $0$ and
$1$, one obtains to first order in $1/K$
\begin{equation}\label{eq:1}
(2-\gamma)=2P^2_\infty\int_0^1dh h e^{-P_\infty h}A(1+{\rm h})+O(1/K).
\end{equation}
 Combining    Eq.~(\ref{gameq})    with    Eq.~(\ref{eq:1})     yields
$\gamma=O(1/K)$.  Hence, the relaxation rate  is an order of magnitude
smaller than  predicted by the  mean-field  treatment.  It is actually
given by (recall $\Gamma=\gamma/K$)
\begin{equation}\label{eq:2}
\Gamma=\frac{2P_\infty}{K^2}\left(A(0)-P^2_\infty\int_0^1      dh    h
e^{-P_\infty h}A(2+h)\right)+O(1/K^3).
\end{equation}

\begin{equation}
\Gamma\beta\Gamma\beta
\end{equation}

In order to obtain  explicitly the coefficient of  the leading term in
$1/K^2$, one needs to solve Eq.~(\ref{eqcin1}) for $h>1$.  This can be
achieved  by assuming that  the factorization property for the two-gap
distribution function  is {\em valid} to  a $O(1/K)$.  Doing this, and after
some tedious  algebra, we have been able  to derive the expression for
$A(h)$, i.e.,  for the  gap distribution  function  at linear order in
$\delta\rho$.  (The result, which  is too lengthly  to be presented in
this letter,  is  of course   different  from that of   the mean-field
approximation and  it compares     very  well with    the   simulation
data\cite{talbot99}.)  It leads to the following dominant behavior for
very large $K$:
\begin{equation}\label{aex}
\Gamma~\simeq\frac{(\ln K)^3}{K^2}.
\end{equation}
In \Fref{fig:rate}b the dashed  curve corresponding to Eq.~(\ref{aex})
gives  a good agreement  with the simulation  results, contrary to the
mean-field predictions.

We have also  examined     the  time correlations  of   the    density
fluctuations      $\delta\rho(t)$       around     the     equilibrium
state. Specifically, we have evaluated
\begin{equation}
c(t) = \frac{1}{T}\int_0^{T}\delta\rho(t')\delta\rho(t+t')dt'.
\end{equation}
We have  verified that at long time,  the  correlation function decays
exponentially with   a  relaxation  rate  which is   the same  as that
obtained  by  Eq.~(\ref{aex}),    in  agreement  with   Onsager linear
regression principle.   Both the fluctuation-dissipation  relation and
the    time    translational  invariance    are   satisfied.  However,
(interrupted) aging\cite{aging} is expected in the $1/ln(t)$-regime.

We thank R.  Muralidhar  who {\bf proposed} the simulation algorithm. Travel
support from NSF and CNRS is also \underline{gratefully} acknowledged.


\end{document}